\documentclass[12pt,tightenlines,nofootinbib]{revtex4}
\usepackage{amssymb}
\usepackage{amsmath}
\usepackage{amsfonts}
\usepackage[stable]{footmisc}
\usepackage{hyperref,blkcntrl,moredefs,relsize,attrib,tikz,booktabs,capt-of,float,varwidth,epsdice,graphicx}

\begin{document}

\title{Emergent deterministic systems\footnote{This article was submitted under the title `God's Dice and Einstein's Solids' to the Spring 2017 FQXi essay contest.}}

\author{Ian T. Durham}
\email[]{idurham@anselm.edu}
\affiliation{Department of Physics, Saint Anselm College, Manchester, NH 03102}
\date{\today}

\begin{abstract}
According to quantum theory, randomness is a fundamental property of the universe yet classical physics is mostly deterministic. In this article I show that it is possible for deterministic systems to arise from random ones and discuss the implications of this for the concept of free will.
\end{abstract}

\maketitle

\begin{quote}
\textit{What does chance ever do for us?} 
\begin{flushright}---William Paley (1743-1805)\end{flushright}
\end{quote}

\section{Classification of process}

What role does chance play in the evolution of the universe? Until the development of quantum mechanics, the general consensus was that what we perceived as chance was really just a manifestation of our lack of complete knowledge of a situation. In a letter to Max Born in December of 1926, Einstein wrote of the new quantum mechanics, ``The theory says a lot, but does not really bring us any closer to the secret of the `old one.' I, at any rate, am convinced that \textit{He} does not throw dice.''~\cite{Einstein:1971aa}. Yet all attempts to develop a deterministic alternative to quantum mechanics have thus far failed. At the most fundamental level, the universe appears to be decidedly random. How is it, then, that the ordered and intentional world of our daily lives arise from this randomness? Intuitively, we tend to think of randomness as being synonymous with unpredictability. But the very nature of the term `unpredictable' implies agency since it implies that an agent is actively making a prediction. I am not interested here in whether or not the universe requires an agent to make sense of it. I'm interested in understanding if it is possible for intentionality to arise naturally from something more fundamental and less ordered.

While we typically think of the universe as being a collection of `things'---particles, fields, baseballs, elephants---it is the processes that these things participate in that make the universe interesting. A process need not require an active agent. A process is simply a change in the state of something.\footnote{It would seem necessary to define `state' here, but given the limitations on length we will leave that for another essay.} This definition of process is similar to the concept of a \textit{test} from operational probabilistic theories (see~\cite{Chiribella:2010aa} for a discussion of such theories). Similar to such theories, then, we can define a \textbf{deterministic} process as being one for which the outcome can be predicted with certainty. To put it another way, a deterministic process is one for which there is only a single possible outcome. A random process must then be an \textit{unpredictable} change. That is, a process for which there is more than one possible outcome is said to be \textbf{random} if all possible outcomes are equally likely to occur. It's important to note that there is a difference between the \textit{likelihood} of making an accurate prediction of the outcome of a process and one's \textit{confidence} in that prediction. Confidence can be quantified as a number between 0 (no confidence) and 1 (perfect confidence). Likelihood is just the probability that a given outcome will occur for a given process. The outcomes of random processes are all equally likely to occur and, in such cases, one's confidence in accurately predicting the correct outcome should be zero.

It is worth noting here the difference between determinism and causality. The concepts are often incorrectly confused. Indeed, D'Ariano, Manessi, and Perinotti have argued that this confusion has led to misinterpretations of the nature of EPR correlations~\cite{dariano14}. Determinism and randomness represent the extremes of predictability. A fully causal theory can have both. To put it another way, in a fully causal theory a random process must still have a cause. Or, as D'Ariano, Manessi, and Perinotti have shown, it is possible to have deterministic processes \textit{without} causality. The difference is that determinism is related to the outcomes of a given process whereas causality is related to the actual occurrence of that process regardless of whether it is deterministic or random.

Of course, many real processes are neither random nor deterministic. There may be more than one possible outcome for a process, but those outcomes may not be equally likely to occur. What do we call such processes? For a two-outcome process whose outcomes have a 51\% and 49\% likelihood of occurrence respectively, one might be tempted to refer to it as `nearly random.' On the other hand, if those same likelihoods were 99\% and 1\% respectively, one might be tempted to say the process was `nearly deterministic.' But what if they were 80\% and 20\% or 60\% and 40\%? At what point do we stop referring to a process as `nearly deterministic' or `nearly random'? We need less arbitrary language. One suggestion would be to refer to such in between cases as `probabilistic.' But this is misleading since we can still assign probabilities to the outcomes of random and deterministic processes; they are no less probabilistic than any other process.

A solution presents itself if we consider the aggregate, long-term behavior of such processes. As an example, consider that casinos set the odds on games of craps---a game that is neither random nor deterministic---under the assumption that they will make money on these games in the long run and (crucially) that the amount of money they will make is reliably predictable within some acceptable range of error. So the process of rolling a pair of dice (which is all that craps is) is at least partially deterministic to a casino. But now consider a game with two outcomes, $A$ and $B$, whose respective probabilities of occurring are 50.5\% and 49.5\%. Could a casino set up a system by which they could, within some range of error, make a long-term profit on this game, even if that profit is very small? Suppose it costs \$100 to play this game and that a player receives \$102 if outcome $B$ occurs but nothing if outcome $A$ occurs. Suppose also that, on average, the casino expects 10,000 people to play this game each year. That means that, on average, they will pay out \$504,900 a year in winnings but keep \$505,000 a year in fees leaving them with \$100 in profit (on average). Though this is ridiculously low, the crucial point is that \textit{it is not zero}. As low as it is, the casinos can still budget for it and, in the long run, can expect to make a profit on it. The fact is that something like this can be done for \textit{any} process that is not random. Non-random processes are always predictable in the aggregate, though the sample size may need to be exceedingly large. Since deterministic processes are perfectly predictable for every occurrence, it makes sense to refer to processes that are predictable only in the aggregate as \textbf{partially deterministic} since they do contain an certain deterministic element to them.

\section{A simple example}

The aforementioned game of craps simply involves betting on the outcome of a roll of a pair of dice. The game is as old as dice themselves and serves as a useful example of how some level of partial determinism can arise from randomness. It also provides a straightforward method for introducing a few additional terms. Those wishing to delve more deeply into this subject are encouraged to dive into Refs.~\cite{Moore:1997aa,Schroeder:2000aa,Ben-Naim:2008aa}.

Consider a fair, six-sided die. As a fair die, it is assumed that upon rolling this die, all of the six outcomes are equally likely. In fact casinos paint the dots on their dice, rather than use the usual divots because the divots are not equally distributed and thus throw off the center-of-mass which changes the long-term probabilities.\footnote{They also routinely replace their dice since the sides of dice can wear unevenly. See Ref.~\cite{Jaynes:1978aa}.} While real dice are never truly random, a so-called `fair die' is considered to be a theoretical ideal and is thus random.

Now consider a roll of two fair dice as in a game of craps. Since they are both fair dice, each outcome on each individual die is equally likely. We are also assuming that we can easily distinguish the dice from one another e.g. perhaps one is blue and one is red. Considered together, then, there are thirty-six possible outcomes---configurations---to a single, simultaneous roll of both. Since we can distinguish between the two dice, if the roll produces a four on the blue die and a three on the red die, this is an entirely different outcome from a three on the blue die and a four on the red one. Each of these configurations is referred to as a \textit{microstate}.

But in craps, as in other games that use a pair of dice, we are often interested in the \textit{sum} of the numbers on the faces. Thus we typically consider the roll of a pair of dice as giving us a number between two and twelve. We call this number the \textit{macrostate}. If we look at each of the thirty-six microstates, we'll see that they can be grouped according to which macrostate they produce. The number of microstates that will produce a given macrostate is known as the \textit{multiplicity} and is given the symbol $\Omega$. But note that the multiplicities of the macrostates for the roll of a pair of fair dice are not all equal. There are, for instance, six different combinations that can produce a roll of seven (I gave two of these six above). On the other hand, there is one and only one way to roll a two or a twelve.

The probability of a given roll (i.e. macrostate) is given by the multiplicity of that roll divided by the \textit{total} multiplicity. So, for example, the probability of rolling a seven is six divided by thirty-six or one-sixth. Conversely, the probability of rolling a two or a twelve is one-thirty-sixth. Table~\ref{tab1} lists the microstates for each macrostate of the pair of dice, giving the multiplicity and probability of each. 
\begin{table}
\begin{center}
\begin{tabular}{p{2.5cm} p{5.5cm} p{1cm} p{2.5cm}}
Macrostate & Microstates & $\Omega$ & Probability \\
\midrule[1pt]
2 & \epsdice[white]{1}\epsdice[black]{1} & 1 & 1/36 \\
3 & \epsdice[white]{1}\epsdice[black]{2}, \epsdice[white]{2}\epsdice[black]{1} & 2 & 2/36=1/18 \\
4 & \epsdice[white]{1}\epsdice[black]{3}, \epsdice[white]{2}\epsdice[black]{2}, \epsdice[white]{3}\epsdice[black]{1} & 3 & 3/36=1/12 \\
5 & \epsdice[white]{1}\epsdice[black]{4}, \epsdice[white]{2}\epsdice[black]{3}, \epsdice[white]{3}\epsdice[black]{2}, \epsdice[white]{4}\epsdice[black]{1} & 4 & 4/36=1/9 \\
6 & \epsdice[white]{1}\epsdice[black]{5}, \epsdice[white]{2}\epsdice[black]{4}, \epsdice[white]{3}\epsdice[black]{3}, \epsdice[white]{4}\epsdice[black]{2}, \epsdice[white]{5}\epsdice[black]{1} & 5 & 5/36 \\
7 & \epsdice[white]{1}\epsdice[black]{6}, \epsdice[white]{2}\epsdice[black]{5}, \epsdice[white]{3}\epsdice[black]{4}, \epsdice[white]{4}\epsdice[black]{3}, \epsdice[white]{5}\epsdice[black]{2}, \epsdice[white]{6}\epsdice[black]{1} & 6 & 6/36=1/6 \\
8 & \epsdice[white]{2}\epsdice[black]{6}, \epsdice[white]{3}\epsdice[black]{5}, \epsdice[white]{4}\epsdice[black]{4}, \epsdice[white]{5}\epsdice[black]{3}, \epsdice[white]{6}\epsdice[black]{2} & 5 & 5/36 \\
9 & \epsdice[white]{3}\epsdice[black]{6}, \epsdice[white]{4}\epsdice[black]{5}, \epsdice[white]{5}\epsdice[black]{4}, \epsdice[white]{6}\epsdice[black]{3} & 4 & 4/36=1/9 \\
10 & \epsdice[white]{4}\epsdice[black]{6}, \epsdice[white]{5}\epsdice[black]{5}, \epsdice[white]{6}\epsdice[black]{4} & 3 & 3/36=1/12 \\
11 & \epsdice[white]{5}\epsdice[black]{6}, \epsdice[white]{6}\epsdice[black]{5} & 2 & 2/36=1/18 \\
12 & \epsdice[white]{6}\epsdice[black]{6} & 1 & 1/36 \\
\midrule[1pt]
& \hfill Total: & 36 & 1
\end{tabular}
\end{center}
\caption{\label{tab1} This table lists the microstates for each macrostate for a roll of a pair of six-sided dice. The multiplicity, $\Omega$, is the total number of microstates.}
\end{table}
Though we think of this as a single roll of a pair dice, it is really two simultaneous rolls of individual dice. Each of these individual rolls is a random process yet when they are considered together as a single roll that single roll of the pair is partially deterministic. As should be clear from Table~\ref{tab1} this behavior is not physical in the sense that the probabilities of the individual macrostates are due to the combinatorics of the problem or what what might call `mindless' mathematics. So a pithy counter to Einstein's objection might be that a dice-throwing God still produces a partially predictable result.

There is one objection to this example that is worth considering. The numbers on the dice are entirely arbitrary. That is, we could have instead painted six different animals on the faces of each die. In this case we might find it hard-pressed to identify any distinctive macrostates other than pairs and we could eliminate the pairs by painting different animals on each die. Thus it seems as if the macrostates used in typical die rolls are entirely arbitrary in the sense that their relative import is based on a meaning that we \textit{assign} to them. The labelling of the sides of the dice is not a fundamental property of the dice themselves. We can get around this problem and improve on our odds by perhaps ironically considering a model proposed by Einstein nineteen years before his comment to Born.

\section{Einstein solids}

In 1907 Einstein proposed a model of solids as sets of quantum oscillators. That is, each atom in such a solid is modeled in such a way that it is allowed to oscillate in any one of three independent directions. Thus a solid having $N$ oscillators would consist of $N/3$ atoms. Crucially, since the oscillators are quantum, they can only hold discrete amounts of energy. So, for instance, suppose that we have an overly simplified Einstein solid containing just a single atom and thus three oscillators. If we supply that solid with a single discrete unit of energy, that energy unit could be absorbed by any one of the three oscillators, but cannot be further subdivided and shared among them. It is assumed that the process of absorption is random. In other words, for a macrostate consisting of a single unit of energy, there are three equally likely microstates, i.e. $\Omega = 3$. In general, for an Einstein solid with $N$ oscillators and $q$ units of energy, the multiplicity is
\begin{equation}
\Omega(N,q) = \left(\begin{array}{c}q+N-1 \\ q\end{array}\right)=\frac{(q+N-1)!}{q!(N-1)!}.
\label{multiplicity}
\end{equation}

Now consider two Einstein solids that are weakly thermally coupled and approximately isolated from the rest of the universe. By weakly thermally coupled, I mean that the exchange of thermal energy between them is much slower than the exchange of thermal energy among the atoms within each solid. This means that over sufficiently short time scales the energies of the individual solids remain essentially fixed. Thus we can refer to the macrostate of the isolated two-solid system as being specified by the individual fixed values of internal energy. (For a further discussion, see Ref.~\cite{Schroeder:2000aa}.) Let's begin by considering a simple (albeit unrealistic) system. Suppose each of our two solids has three oscillators, i.e. $N_A = N_B = 3$, and the system has a total of six units of energy that can be divvied up between the oscillators. Suppose that we put all six of these units of energy into solid $B$. That means that there is only one possible configuration for the oscillators in solid $A$---they all contain zero energy. Conversely, there are twenty-eight configurations for the oscillators in solid $B$ according to~(\ref{multiplicity}). The total number of configurations for the system as a whole is just the product of the two and thus is also twenty-eight.

Suppose that we instead put a single unit of energy into solid $A$ with the rest going to solid $B$. In this case, there are three possible configurations for solid $A$ since the single unit of energy we've supplied to it could be in any one of the three oscillators. The five remaining units of energy can be distributed in any one of twenty-one ways within solid $B$. But now the \textit{total} number of configurations for the system is $3\cdot21=63$. Table II summarizes the energy distribution and corresponding multiplicity for this simple system and Fig. I shows a smoothed plot of the total multiplicity, $\Omega_{\rm{tot}}$ as a function of the energy $q_A$ contained in solid $A$.
\begin{center}
\begin{table}[h!]
\begin{varwidth}[b]{0.49\textwidth}
  \centering
  \begin{tabular}{ p{0.75cm} p{1cm} p{0.75cm} p{1cm} c}
  $q_A$ & $\Omega_A$ & $q_B$ & $\Omega_B$ & $\Omega_{\rm{tot}}=\Omega_A\Omega_B$ \\
  \midrule[1pt]
  0 & 1 & 6 & 28 & 28 \\
  1 & 3 & 5 & 21 & 63 \\
  2 & 6 & 4 & 15 & 90 \\
  3 & 10 & 3 & 10 & 100 \\
  4 & 15 & 2 & 6 & 90 \\
  5 & 21 & 1 & 3 & 63 \\
  6 & 28 & 0 & 1 & 28
  \end{tabular}
  \caption{This table shows the distribution of six units of energy among two Einstein solids, each with three oscillators.}
  \label{table:tab2}
\end{varwidth}
\hfill
\begin{minipage}[b]{0.49\textwidth}
  \begin{tikzpicture}[align=left]
  \node at (2,1.5) {
    \includegraphics[width=5cm]{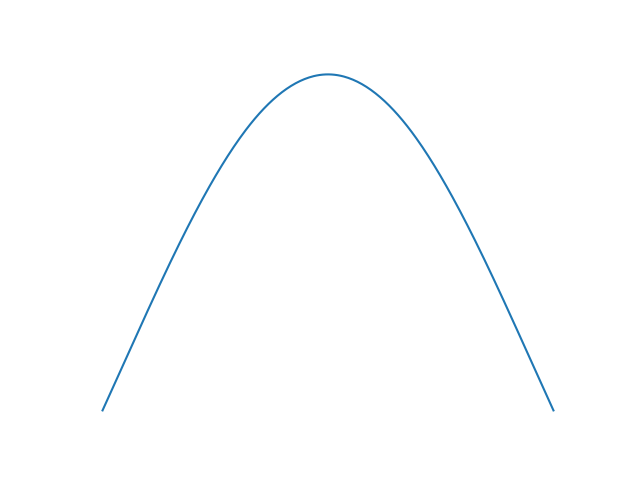}
  };
  \draw[thick,-latex] (0,0) -- (4.5,0);
  \draw[thick,-latex] (0,0) -- (0,3.5);
  \node[left] at (0,3.25) {$\Omega_{\rm{tot}}$};
  \node[below] at (4.5,0) {$q_A$};
  \node at (4,3) {\scriptsize{$N_A=N_B=3$} \\ \scriptsize{$q=6$}};
  \end{tikzpicture}
  \captionof{figure}{This shows a smoothed plot of $\Omega_{\rm{tot}}$ as a function of $q_A$ for the data from Table I.}
  \label{figure:smallsysmult}
\end{minipage}
\end{table}
\end{center}
This tells us that the states for which the energy is more evenly balanced between the two solids are more likely to occur because there are more possible ways to distribute the energy in such cases. This is analogous to the example given in the previous section involving a pair of fair dice. There is no intentionality on the part of the system. In addition, the system is considered to be isolated from the rest of the universe and thus there is no environment driving these results. They are simply due to combinatorics. Each individual microstate of the combined system is assumed to be equally probable and thus the process of reaching one of these microstates from any other is completely random. It just happens that more of those microstates correspond to configurations in which the energy is more evenly divided between the two solids. Thus the system can undergo random fluctuations about the mean and still be more likely to be found in a microstate in which the energy is roughly equally divided between the two solids.

\begin{figure}
\begin{center}
\begin{tabular}{ccc}
\begin{tikzpicture}
\node at (2,1.5) {
  \includegraphics[width=5cm]{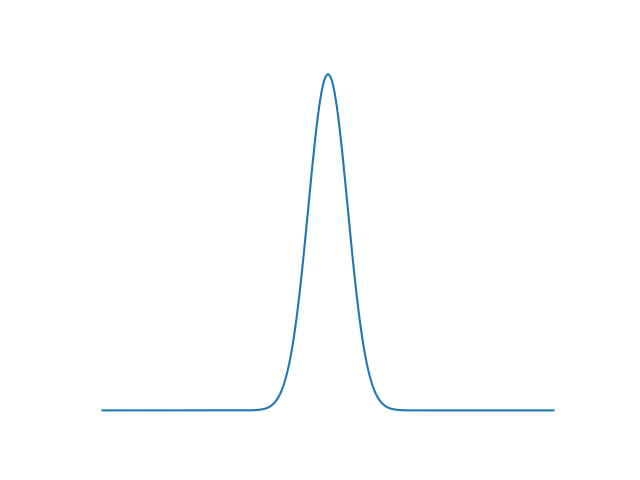}
};
\draw[thick,-latex] (0,0) -- (4.5,0);
\draw[thick,-latex] (0,0) -- (0,3.5);
\node[left] at (0,3.25) {$\Omega_{\rm{tot}}$}; 
\node[below] at (4.5,0) {$q_A$};
\node at (2,3.1) {\scriptsize{$N,q\approx$ a few hundred}};
\end{tikzpicture}
& $\quad$ &
\begin{tikzpicture}
\node at (2,1.5) {
  \includegraphics[width=5cm]{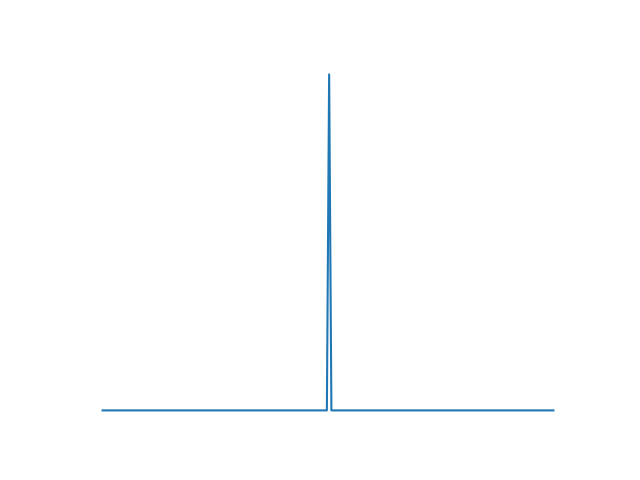}
};
\draw[thick,-latex] (0,0) -- (4.5,0);
\draw[thick,-latex] (0,0) -- (0,3.5);
\node[left] at (0,3.25) {$\Omega_{\rm{tot}}$}; 
\node[below] at (4.5,0) {$q_A$};
\node at (2,3.1) {\scriptsize{$N,q\approx$ a few thousand}};
\end{tikzpicture} \\
& \\
(a) & & (b)
\end{tabular}
\end{center}
\caption{\label{largesysmult} (a) This shows a plot of $\Omega_{\rm{tot}}$ as a function of $q_A$ for a pair of Einstein solids when $N,q\approx$ a few hundred. (b) This is the resultant plot for the solids when $N,q\approx$ a few thousand.}
\end{figure}

But consider now what happens when we begin to scale the system up to more realistic sizes. Fig.~\ref{largesysmult}a shows a plot of the total multiplicity of the system as a function of the energy in solid $A$ when the total number of oscillators and the total number of energy units is a few hundred. 
Fig.~\ref{largesysmult}b shows a plot of the same function when the number of oscillators and energy units is a few thousand. The larger our Einstein solids become, the narrower the peak of the multiplicity function. For realistic Einstein solids, the peak is so narrow that only a tiny fraction of microstates have a reasonable probability of occurring. That is, random fluctuations away from equilibrium are entirely unmeasurable.

It is important to keep in mind that all we have done in Fig.~\ref{largesysmult} is to scale up the Einstein solids. Each individual microstate remains equally probable and thus the underlying process of moving from one microstate to another is entirely random. Yet, as the system grows larger and larger, it is increasingly likely to be found in only a very small number of microstates. This means that we can make \textit{highly} accurate predictions of which microstates will occur given some initial input data. This is a dramatically scaled up analogy to a game of craps. Though the fluctuations taking the system from one microstate to another are entirely random, the \textit{macrostate} is very nearly deterministic. And though I have used a decidedly physical example, the result is simply a consequence of the combinatorial behavior of very large numbers. Thus we have a situation in which a nearly deterministic process can arise from random processes due solely to something that is almost entirely mathematical. In addition, unlike the situation with the dice, we are not arbitrarily assigning meaning to the microstates and macrostates.

An obvious question is whether or not it is possible to achieve \textit{perfect} determinism with a pair of Einstein solids. Certainly in the limit that $N,q\to\infty$ the narrowness of the multiplicity peak in our example becomes asymptotically thin. The limit in which a system becomes large enough that random fluctuations away from equilibrium become unmeasurable is known as the \textit{thermodynamic limit}. In other words, at some point our partially deterministic system becomes indistinguishable from a fully deterministic one. Where that transition occurs may depend on a host of factors, but the reason it is referred to as the thermodynamic limit is precisely because it is where conventional thermodynamic methods of analysis---which are deterministic!---become the most useful way to understand the behavior of a pair of solids that are in thermal contact with one another.

Of course, this is just a single example from one area of physics but it serves to show that near-perfectly deterministic macroscopic processes \textit{can} arise from a very large number of random microscopic processes due to the `mindless' behavior of mathematics. 

\section{Implications for free will}

Let us consider a toy universe in which all microscopic processes are random and thus equally probable. The only physical constraints that we will place on this toy universe are to limit the outcomes of each microscopic process to being finite in number and to require that these outcomes be distinguishable from one another. Macroscopic processes in such a toy universe would have varying levels of determinism based on the combinatorics and the nature of the processes themselves. For example, the microstates and macrostates of a pair of six-sided dice are different from the microstates and macrostates of one six-sided die and one eight-sided die. Thus the nature of the dice dictate which processes are allowed in each case (e.g. a roll of fourteen is not possible with a pair of six-sided dice). For our toy universe, we can think of any constraints as being dictated by the initial physical conditions of the universe itself.

It is worth asking, then, what it would mean for a hypothetical `being' in such a universe to have free will. Free will is generally viewed as one's ability to freely choose between different courses of action. This requires, however, that when presented with a choice, an agent can reliably predict the outcome of some process. If I am, for instance, faced with the choice of carrots or broccoli as a vegetable side for my dinner, the essence of free will is that, free of unpredictable external factors, if I choose to have carrots I can have confidence that I will actually have carrots with my dinner, i.e. the carrots won't randomly and inexplicably turn into a potato the moment they touch my plate. The crucial but subtle difference here is that my choice in this example is between two \textit{different processes}---the process of physically taking carrots from my refrigerator or the process of physically taking broccoli from my refrigerator---rather than two different outcomes of a \textit{single} process. So once I have chosen to carry out one or the other of these processes, I can have confidence that the multiplicity of one outcome of my chosen process is so much greater than the multiplicity of any other outcome that my desired result will actually occur, i.e. the probability of the most likely macrostate \textit{not} occurring is utterly unmeasurable.

Of course, any beings in our hypothetical toy universe are unequivocally \textit{part} of that universe and thus an amalgam of random processes themselves. If the deterministic macroscopic processes arise from microscopic random ones solely due to the combinatorics of a large number of such microscopic processes, then it is worth asking if free will really does exist. This is certainly a fair question, but misses the broader point. Regardless of what happens at the most fundamental level, the concept of free will is meant to be applied to sentient beings (which are inherently \textit{not} fundamental) making conscious choices about the macrostates of large-scale systems. As sentient beings we expect that free will entails our ability to freely make a choice with the confidence that a specific outcome of our chosen process really does occur with a high degree of probability. For that to happen certain processes must be at least partially deterministic if not fully so.

This brings up an important distinction. There are really different \textit{levels} of processes. We can refer to a process associated with a macrostate as a \textit{macroprocess}. The constituent processes of a macroprocess would then be \textit{microprocesses}. The macroprocess of simultaneously rolling a single pair of dice is composed of two microprocesses---the independent rolls of two individual dice. So the terminology refers to the level of the system and not necessarily the size of the system or its constituents. The act of me pulling carrots out of a bin in my refrigerator is a macroprocess that actually consists of trillions of microprocesses involving the neurons in my brain, the electrical signals in my neurological system, the mechanical motion of the refrigerator parts, etc. These in turn are all made up of further constituent processes all the way down to the processes involving the fundamental particles and fields that constitute the material foundation of the entire system.

Free will thus generally involve choices about \textit{macroprocesses} with varying degrees of confidence. I may be highly confident that the carrots in my refrigerator won't spontaneously turn into potatoes, but I'm a tad less confident that inserting the key into the ignition of my car will turn the car on. Certainly I expect it to turn on \textit{most} of the time, but it is entirely plausible that something could go wrong and it won't turn on. I'm even less confident when I approach an unfamiliar intersection and don't know which way to go. Depending on the situation, my choice could essentially be entirely random. The key point here is that if \textit{all} macroprocesses were entirely random, we wouldn't even have the \textit{illusion} of free will because our choices would be meaningless since they would be based entirely on guesses. So free will requires that most \textit{macroprocesses} be at least partially deterministic. But, crucially, \textit{microprocesses} can still be random since their combinatorial behavior can lead to partially deterministic macroprocesses like the rolling of dice or the equilibrium state of two solids in thermal contact.

\section{Boundary Conditions}

There's one final objection to this line of argument that should be addressed. It's clear that the emergence of determinism and free will in this model is not solely due to the combinatorics alone. After all, the mathematics refers to something physical. As I said before, the behavior of a six-sided die is different from the behavior of an eight-sided die. So at the most fundamental level there has to be something non-mathematical in order to distinguish, for example, a quark from a lepton or even the number one from the number two. But it is worth asking if the combinatorics itself can produce additional boundary conditions on the system that then further constrain its evolution. In other words, is it possible for a system's own internal combinatorics to change the probabilities of future macrostates?

In the simple example using dice, no matter how many times we roll them, the combinatorics alone will not change the probabilities of the macrostates. Certainly the dice could wear down unevenly over time, but this is an external effect. But consider a pair of Einstein solids in thermal contact as I described in the previous section. A microprocess for such a system is the shifting of an energy unit from one oscillator to another. This microprocess is fundamentally random. If we introduce a large number of energy units to such a system and assume it has a large number of oscillators, regardless of how those energy units are initially distributed, over time the system will find itself limited to just a few possible microstates. Crucially, these random microprocesses don't suddenly cease to occur when the system reaches equilibrium. Energy continues to be passed around while the underlying microprocesses remain random, yet fluctuations away from equilibrium eventually become unmeasurable. This is simply because a few microstates near equilibrium have an enormously higher probability of occurring than all the other microstates. In this sense, the macrostate corresponding to equilibrium has imposed a boundary condition on future \textit{macroprocesses} purely through combinatorics. So while the evolution toward equilibrium has no effect on the underlying microprocesses which are presumably fixed by the inescapable laws of physics, it \textit{does} have an effect on the future evolution of the aggregate macroprocesses for entirely combinatorial reasons.

So while free will does not allow us to alter the laws of physics, it does act as an introduced boundary condition that can allow for a certain amount of environmental forcing on the macro-level and it seems that it is at least \textit{possible} for that to happen for purely combinatorial reasons.

\section{Conclusion}

There is little in this article that is actually speculative. Admittedly I am considering highly simplified systems here, but they at least demonstrate that it is possible for something ordered and intentional to arise from the aggregate behavior of a collection of random processes with no external forcing, i.e. due solely to combinatorics. The behavior of such systems also suggests that it is entirely possible for free will to emerge from something far less ordered. In fact both Eddington and Compton argued that the randomness of quantum mechanics was a \textit{necessary condition} for free will~\cite{Eddington:1928aa,Compton:1935aa}. On the other hand, Lloyd has argued that even deterministic systems can't predict the results of their decision-making process ahead of time~\cite{Lloyd:2012aa}. Is free will just an illusion? Does it require randomness or does it require determinism? The answers to these questions undoubtedly lie in a deeper understanding of the transition from quantum systems to classical ones. In this essay I have shown that the seeds of such an understanding might be found in simple combinatorics. The mindless laws of mathematics might just be what allows the universe to evolve intentionality. At the very least, it is worth a deeper look.

\begin{acknowledgements}
I would like to thank Irene Antonenko for pressing me on the language of partial determinism. I stand by my use of the term, but Irene's comments helped me to clarify why I prefer it. Plus it made for a great after-dinner discussion that was enhanced by good dessert and good wine. I additionally thank our spouses and children who cleaned up around us.
\end{acknowledgements} 

\newpage

\bibliographystyle{plain}
\bibliography{FQXi6.bib}

\end{document}